\documentclass[universe,article,moreauthors,accept]{Definitions/mdpi} 
\usepackage{comment} 
\usepackage{graphicx}
\newcommand{\rthis}[1]{\textcolor{black}{#1}}

\firstpage{1} 
\pubvolume{1}
\issuenum{1}
\articlenumber{0}
\pubyear{2026}
\copyrightyear{2026}
\datereceived{ } 
\daterevised{ } 
\dateaccepted{ } 
\datepublished{ } 

\Title{Determination of neutron star radius from pulse profile modeling using profile likelihood}

\Author{Vyaas Ramakrishnan \orcidA{},  Shantanu Desai* \orcidB{}}

\AuthorNames{Vyaas Ramakrishnan, Shantanu Desai}

\address{Department of Physics, Indian Institute of Technology,  Hyderabad, Kandi, Telangana 502284, India}

\corres{Correspondence: shntn05@gmail.com; }

\abstract{In recent years,  NICER data have been extensively used to determine neutron star \rthis{masses} and radii  using pulse profile modeling. Pulse profile modeling is implemented with the {\tt X-PSI} package and the best-fit parameters are typically obtained using Bayesian inference. Using simulated data, we demonstrate the first ever application of frequentist inference to determine the neutron star radius, where the nuisance parameters are treated  using profile likelihood. We find that the profile likelihood technique can recover the true radius to  $< 1\sigma$. The uncertainty in the estimated radius is also comparable to that obtained  from  Bayesian analysis while being computationally much faster. Therefore, this work serves as a proof-of-principle application of frequentist inference to \rthis{estimate} the neutron star radius using pulse profile modeling and complements the Bayesian inference technique currently used. We have also made our analysis codes for frequentist inference using {\tt X-PSI} publicly available.}

\keyword{Neutron Stars; Profile Likelihood; Bayesian Inference} 

\begin{document}

\section{Introduction}

The Neutron Star Interior Composition Explorer
(NICER) is an X-ray satellite that has facilitated precise measurements of the neutron star mass and radius through its timing precision since its launch in 2017~\cite{NICER}. The mass and radius have been determined through detailed analysis of the associated X-ray emission through pulse profile modeling for  a large number of neutron stars~\cite{Watts2016,Gonzalez26}. The mass and radius measurements can provide stringent constraints on the neutron star equation of state~\cite{Ozel}.
The analysis of the X-ray pulse  profiles is usually performed  using the X-ray Pulse Simulation and Inference package (X-PSI)~\cite{XPSI}. This package is a ray-tracing code that models the phase- and energy-dependent emission from hot spots and the bulk of the rotating  neutron star surface. This package enables simultaneous modeling of hot-spot geometry and atmospheric emission, while incorporating both special and general relativistic effects within the oblate Schwarzschild framework. More details of the physics inputs used in X-PSI are described elsewhere~\cite{Morsink14,Bogdanov}.
This package provides parameter estimates for neutron star mass, radius, and emission geometry using Bayesian inference~\cite{Trotta}. The sampling of posteriors required for Bayesian inference is performed using the {\tt MultiNest}~\cite{multinest} and {\tt UltraNest}~\cite{Ultranest} software packages, both based on nested sampling. The {\tt X-PSI} package has been widely used in a number of works~\cite{Gonzalez26,Salmi,Dittman,Mauviard} (and also references therein).

However, discrepancies in the inferred radius have been found among different studies  in the literature using this package. For instance, Ref.~\cite{Salmi} obtained an equatorial radius of   $12.49^{+1.28}_{-0.88}$ km for PSR J0740+6620, whereas a value of $12.76^{+1.49}_{-1.02}$ km was obtained for the same source in Ref.~\cite{Dittman}. The discrepancy was attributed to a combination of multiple factors: the choice  of priors, the posterior sampler used, and a lack of convergence due to improper settings used in the samplers~\cite{Mariska,Salmi}. Similar differences in results from nested samplers have also been found in the posteriors obtained from merging compact binaries~\cite{Bilby}.

To test this conjecture, the parameter-estimation accuracy of {\tt MultiNest} was compared with that of another sampler, namely {\tt UltraNest}, with  synthetic (simulated) data, that mimic the real NICER data of  PSR J0740+6620~\cite{Mariska} (H25, hereafter). {\tt UltraNest} is based on the Sliced Sampling Technique~\cite{SS} and is known to be more robust than {\tt MultiNest}, although it is much slower. Using numerical experiments, it was shown that both samplers produced unbiased results for parameter estimation~\cite{Mariska}. Subsequently, it was also shown that both samplers produce consistent credible intervals on real data from PSR J0740+6620. Nevertheless,  some concerns about the convergence of {\tt MultiNest}
were raised in H25. This work found that convergence was challenging  due to the flat likelihood surface for radii $> 11$ km which made it difficult to constrain the upper limit. Furthermore, as the radius increases, additional degeneracies arise. They also cautioned that higher sampling settings were needed for real data  to fully explore the parameter space. Such tests require more computational resources and hence were not carried out in H25. Furthermore, another caveat with Bayesian parameter inference is that when the  prior ranges for the free parameters are broad, marginalization needed for parameter estimation can be affected by  volume effects,  which could  introduce a non-negligible bias into the parameter estimates~\cite{Adria}.

In order to alleviate the aforementioned concerns regarding Bayesian parameter estimation in {\tt X-PSI}, we demonstrate how one can use frequentist techniques based on profile likelihood to infer the neutron star radius. 
Although Bayesian statistics has been the industry standard in astrophysics and cosmology for the past two decades, in recent years there has been a resurgence of interest in the application of profile likelihood  in parameter estimation~\cite{Herold,Campeti,Karwal,Barua25,Herold24,Vyaas}. We note, however that particle physics and  particle astrophysics still mostly use frequentist inference for parameter estimation~\cite{PDG}. Frequentist parameter estimation does not require the specification of  priors or entail the use of a  sampler, and hence one does not need to worry about convergence issues.
Therefore, frequentist inference should be immune to the inconsistencies found in  Bayesian determinations of the neutron star radius. However, frequentist inference techniques also make implicit assumptions~\cite{Herold24}, and the minimization algorithms implemented could become trapped in local minima, when dealing with a large multidimensional parameter space.
Therefore, given the importance of results from X-PSI for fundamental physics, it is incumbent upon us to compare the results obtained using Bayesian and frequentist inference, which is the goal of this work.

The manuscript is structured as follows. We discuss our analysis in Sect.~\ref{sec:analysis}, results in Sect.~\ref{sec:results}, and conclude in Sect.~\ref{sec:conclusions}.


\section{Analysis} 
\label{sec:analysis}

We describe the methodology used  in both the standard Bayesian and frequentist inference pipelines to compare and contrast the estimated parameter values along with their associated uncertainties.

All analyses were performed using the {\tt X-PSI} pulse profile modeling framework. 
The Bayesian and frequentist pipelines share an identical likelihood implementation and astrophysical model configuration, thereby ensuring a consistent basis for 
comparison between the two methodologies. We now describe each of them below.

\subsection{Bayesian analysis}
For Bayesian analysis, we follow the same settings as in H25.  We provide a brief description of Bayesian parameter inference below. More details can be found in ~\cite{Trotta08}. The posterior probability  $P(\theta|d)$ for the dataset $d$ can be written as:   
\begin{equation}
P(\theta|d) \propto  \mathcal{L} (d,\theta) Pr(\theta), 
\end{equation}
where $\mathcal{L}$ is the likelihood used, $\theta$ denotes the parameter vector containing the free parameters used in the model, and Pr($\theta$) denotes the prior for each of the free parameters in $\theta$.
{\tt X-PSI} uses a Poisson likelihood, whose details can be found elsewhere~\cite{Riley21}. {\tt X-PSI} allows for up to 16 free parameters, as  tabulated in Table 1 of ~\cite{Riley21}.
For this work, the  Bayesian posterior was sampled using {\tt MultiNest} interfaced with {\tt X-PSI}. The adopted pulse profile model was the ST-U configuration, consisting of two single-temperature, simply connected circular hot regions with independent parameters. The neutron star atmosphere was modeled using the NSX fully ionized hydrogen atmosphere model~\cite{Ho01,Ho09}, while photon propagation was calculated using general-relativistic ray tracing in the oblate Schwarzschild spacetime approximation.

\subsection{Frequentist analysis}
Frequentist parameter estimation was performed using a profile likelihood framework implemented directly on top of the {\tt X-PSI} likelihood evaluation module. No modifications were made to the underlying physical model. All  settings were kept unchanged from H25, ensuring that differences between the Bayesian and Frequentist analyses arose solely from the inference methodology. We provide  a  brief outline of parameter estimation using profile likelihood.
More details  and applications to astrophysics and cosmology can be found in ~\cite{Vyaas,Herold24} (and references therein).

Let us consider a parameter estimation problem  where the parameter vector ($\theta$) consists of two parameters: $\theta \equiv \{\phi,\alpha\}$.  Here, $\phi$ denotes the  parameter of interest and $\alpha$ the nuisance parameter. In frequentist statistics  the profile likelihood is calculated  by maximizing the  likelihood $\mathcal{L}(\phi,\alpha)$ with respect to $\alpha$:
\begin{equation}
\mathcal{L}(\phi)= \max_{\alpha} \mathcal{L}(\phi,\alpha) .
\label{eq:pL}
\end{equation}
This method is sometimes referred to as graphical profile likelihood in the literature~\cite{Herold24}.  The parameter estimate for $\phi$
can be obtained from the distribution of $\mathcal{L}(\phi)$. For computational simplicity, we define  $\chi^2(\phi) \equiv -2\ln  \mathcal{L}(\phi)$, and the frequentist confidence intervals are constructed from  $\Delta \chi^2 (\phi) = \chi^2 (\phi) - \chi^2_{min}$,
where $\chi^2_{min}$ is the global minimum of $\chi^2(\phi)$. To obtain confidence intervals, we  use Wilks' theorem,  which states that  $\Delta\chi^2$ follows a $\chi^2$ distribution  with one degree of freedom~\cite{Wilks1938,Herold24}.

For our particular problem, the parameter of interest, the equatorial radius, $R_{eq}$, was held fixed while the likelihood was maximized over the remaining 12 free nuisance parameters. These nuisance parameters  include the neutron star mass, source distance, cosine of the observer inclination, primary and secondary hotspot phase shifts, colatitudes, angular radii, temperatures, and the interstellar hydrogen column density. Hence, we obtain a likelihood profile for the equatorial radius, where each point on the profile represents the maximum likelihood obtained after optimizing over all other free model parameters. The optimization was carried out using the Nelder--Mead simplex algorithm~\cite{NR} as implemented in \texttt{scipy.optimize.minimize}.

\section{Results}
\label{sec:results}

The Bayesian and frequentist inference procedures were applied to the same synthetic NICER and XMM-Newton datasets discussed in H25. \rthis{H25 provided 10 synthetic  PSR J0740+6620-like NICER and XMM-Newton datasets. For this analysis, we use the dataset in the first column in Table 1 of H25,  labelled (\textit{syntX1}).
In Appendix A, we perform a similar test with another such synthetic dataset.}

\begin{figure}[H] 
    \centering      
    \includegraphics[width=0.85\linewidth]{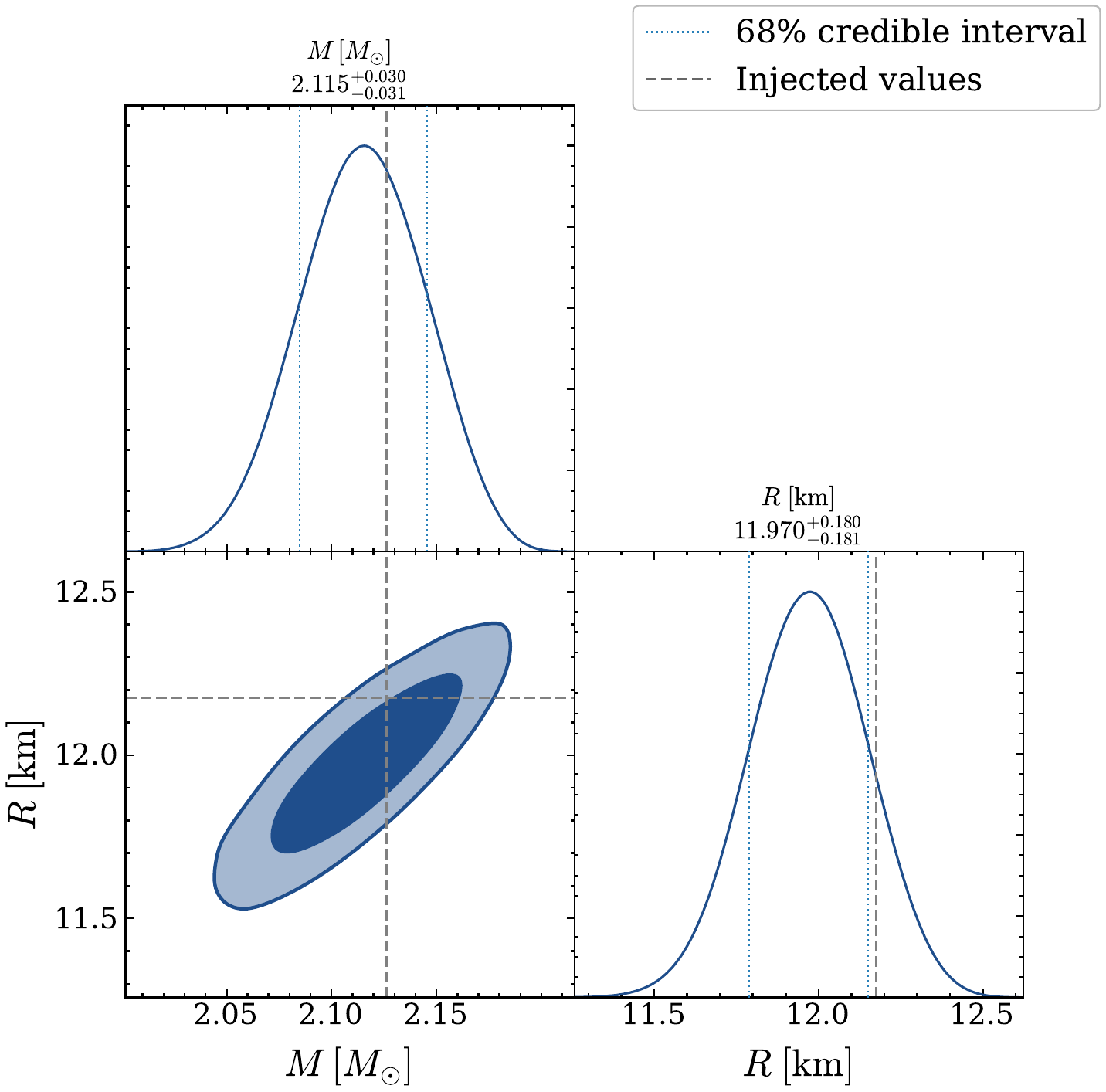}
    \caption{ Marginalized (68\% and 95\%) credible intervals  (obtained using the {\tt MultiNest} sampler) for neutron star mass $M\,[M_{\odot}]$ and equatorial radius $R\,[\mathrm{km}]$, along with the 1-D 68\% credible intervals, inferred from X-ray pulse profile modeling using the synthetic data subset, \textit{syntX1}. The gray dashed line indicates the injected (true) parameter values. The corner plots were obtained using {\tt getdist} package~\cite{getdist}.}
    \label{fig1}
\end{figure}

\begin{figure}[H] 
    \centering      
    \includegraphics[width=0.85\linewidth]{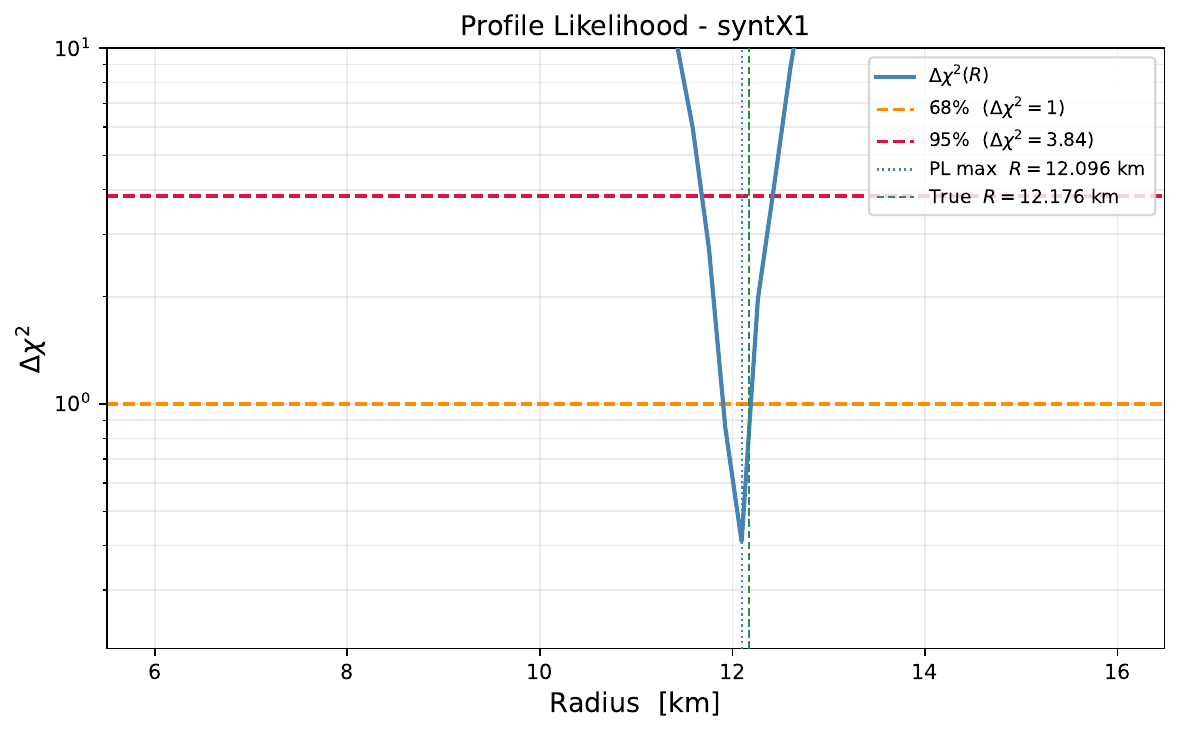}
    \caption{$\Delta\chi^2$   as a function of the neutron star radius, inferred from X-ray pulse profile modeling using the synthetic data subset \textit{syntX1}, where $\Delta\chi^2 = -2[\ln\mathcal{L}_{\mathrm{profile}} - \ln\mathcal{L}_{\mathrm{max}}]$. \rthis{The profile likelihood ($\mathcal{L}_{\mathrm{profile}}$) is defined in Eq.~\ref{eq:pL} and $\mathcal{L}_{\mathrm{max}}$ is the global maximum of the likelihood.}
    The horizontal dashed lines mark the 68\% ($\Delta\chi^2 = 1$) and 95\% ($\Delta\chi^2 = 3.84$) confidence thresholds. The dotted green vertical line denotes the true injected radius, while the blue dotted line illustrates the recovered best-fit radius from the profile likelihood analysis.}
    \label{fig2}
\end{figure}

The resulting Bayesian credible intervals  for the mass and radius, together with the frequentist profile likelihood are shown in Figures~\ref{fig1} and \ref{fig2}, respectively. We now present the results for each analysis.

Figure \ref{fig1} shows the 68\% and 95\% credible intervals for two parameters, namely the mass and the radius of the neutron star, obtained from the \texttt{MultiNest} nested-sampling analysis. The joint mass-radius posterior is marginalized over the remaining 11 nuisance parameters. The true injected radius  corresponding to the truth value in the simulated data and specific to \textit{syntX1}, is also overlaid for comparison. The posterior distributions for the neutron star mass and the equatorial radius are well localized around the injected values, indicating successful recovery of the underlying model parameters. Hence, we obtain the 68\% credible intervals for the neutron star mass and equatorial radius.
The 68\% marginalized credible interval for the radius is  $R_{68,bayes} = 11.970^{+0.180}_{-0.181}$ km, whereas the injected  radius is  12.176 km. Therefore, the radius inferred using Bayesian analysis agrees with the injected radius to within $1.1\sigma$.

Figure~\ref{fig2} presents the frequentist profile likelihood as a function of the neutron star equatorial radius, obtained by maximizing the likelihood for each radius value over all other model parameters. We observe a single, well-defined minimum at $R=12.096$ km, in the vicinity of the true injected radius. Furthermore, the injected radius, $R = 12.176$ km, lies close to the profile likelihood estimate, and is consistent with the inferred confidence intervals from the $\Delta\chi^2$ profile. The $68\%$ confidence interval for the radius is $R_{68} = 12.096^{+0.064}_{-0.182}$ km, and the $95\%$ confidence interval for the radius is  $R_{95} = 12.096^{+0.316}_{-0.397}$ km. Therefore, the frequentist estimate for the radius agrees with the injected radius to within $0.4\sigma$, assuming the larger of the two 68\% confidence-interval uncertainties. This significance of $0.4\sigma$ is based on the difference between the frequentist and Bayesian estimates, divided by the uncertainty in the Bayesian estimate.  
This precision is also comparable to that of the Bayesian estimate.

To quantify the computational efficiency of the two inference frameworks, we compare their total computational costs for completing an inference run. For this comparison, we  consider only  the analysis runs using the  low-resolution settings for {\tt X-PSI}, as described in H25~\cite{Mariska}. The frequentist profile likelihood analysis required approximately 9 CPU-core hours, compared with $\approx 80$ wall-clock  hours on a 48-core computing node for the Bayesian analysis. This corresponds to approximately 3,840 CPU-core hours and is of the same order of magnitude as the computational cost reported in H25 (4.8K CPU-core hours). Despite the lower computational cost, the profile likelihood approach yielded parameter constraints that are comparable in precision with the Bayesian estimate. Further, the difference from the injected value is smaller for the frequentist estimate ($0.4 \sigma$) than for the Bayesian estimate ($1.1\sigma$).
Therefore, our results demonstrate that frequentist parameter estimation can complement Bayesian estimates of neutron star radius using pulse profile modeling.

\section{Conclusions}
\label{sec:conclusions}

For this comparison, we present a proof-of-principle determination of the neutron star radius from pulse profile modeling with the {\tt X-PSI} package,  using  a  frequentist profile likelihood framework to complement the Bayesian inference technique  normally used.
Using synthetic data, we demonstrate  that frequentist inference can recover the true neutron star radius to an accuracy of  $< 1\sigma$. Its precision based on synthetic data is comparable to that of  Bayesian analysis, while requiring substantially less computational time than the {\tt MultiNest}-based Bayesian inference used in {\tt X-PSI}. We caution, however, that all our tests are  based only on simulated data using the {\tt X-PSI} package.
This package does not simulate the full NICER observing pipeline or the detector environment.
A non-exhaustive list of systematics that are not included in this package but  affect real NICER observations includes particle-induced background, diffuse X-ray background, unresolved astrophysical sources, time-variable or orbit-dependent background, solar activity and geomagnetic effects, among others~\cite{Rem22}.
We have not yet  applied this frequentist technique to real NICER or XMM-Newton data. Therefore, it remains to be determined whether the conclusions drawn from the application to synthetic data can be extrapolated to real observations.
In future work, we shall investigate this by applying the technique to real pulsar data from  NICER, XMM-Newton, and other observatories.

In the spirit of open science, we have made our profile likelihood-based modification to {\tt X-PSI} publicly available, which can be found at \url{https://github.com/vyaas3305/xpsi-frequentist-inference}.

\appendixtitles{Yes} 
\appendixstart
\appendix
\section{Test with additional synthetic dataset}
\begin{figure}[H] 
    \centering      
    \includegraphics[width=0.85\linewidth]{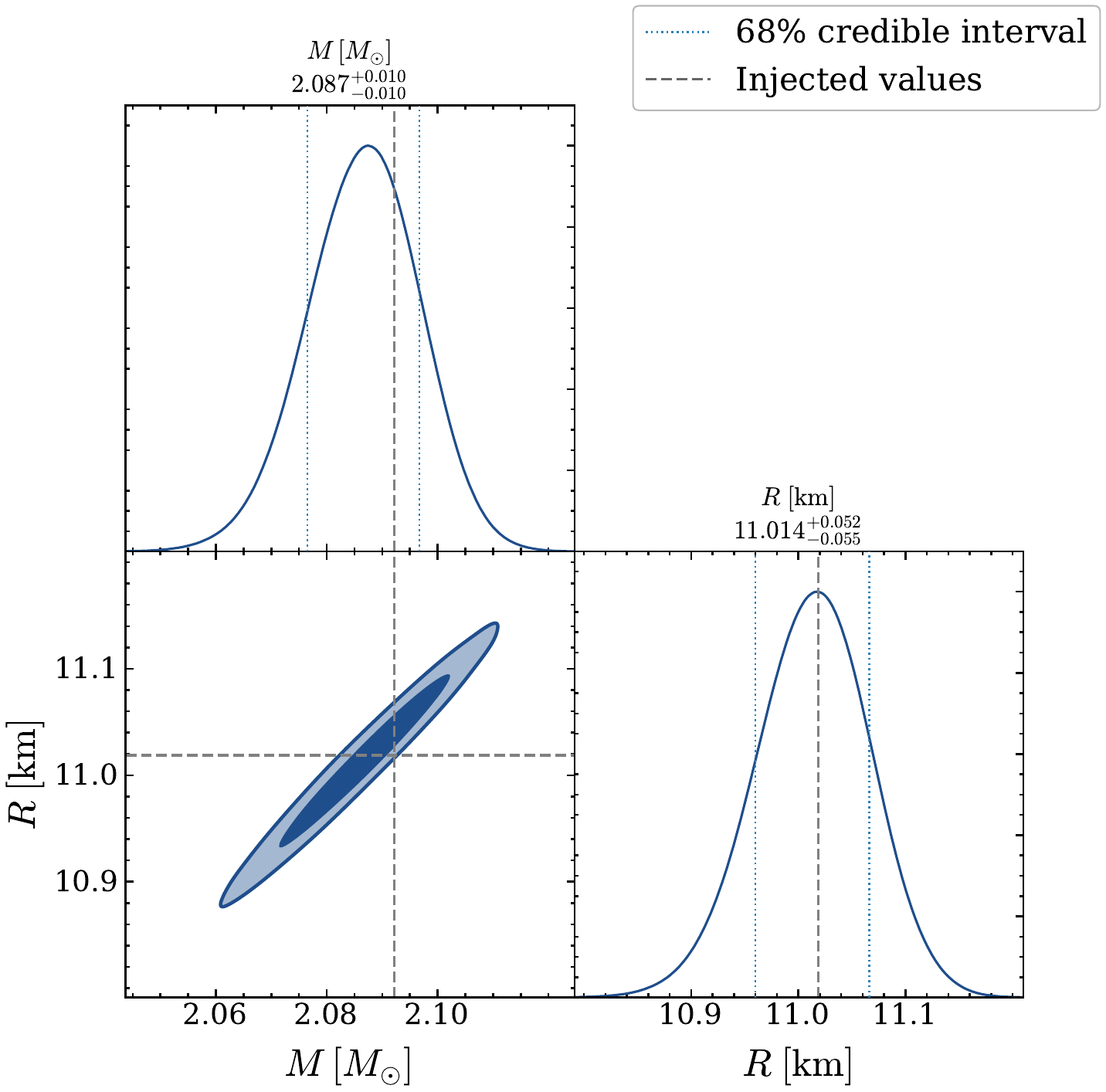}
    \caption{ Marginalized (68\% and 95\%) credible intervals  (obtained using the {\tt MultiNest} sampler) for neutron star mass $M\,[M_{\odot}]$ and equatorial radius $R\,[\mathrm{km}]$, along with the 1-D 68\% credible intervals, inferred from X-ray pulse profile modeling of synthetic data subset, \textit{syntX2}. \rthis{The gray vertical dashed line indicates the injected (true) parameter values. The corner plots were obtained using the {\tt getdist} package.}}
    \label{fig:d2B}
\end{figure}

\begin{figure}[H] 
    \centering      
    \includegraphics[width=0.85\linewidth]{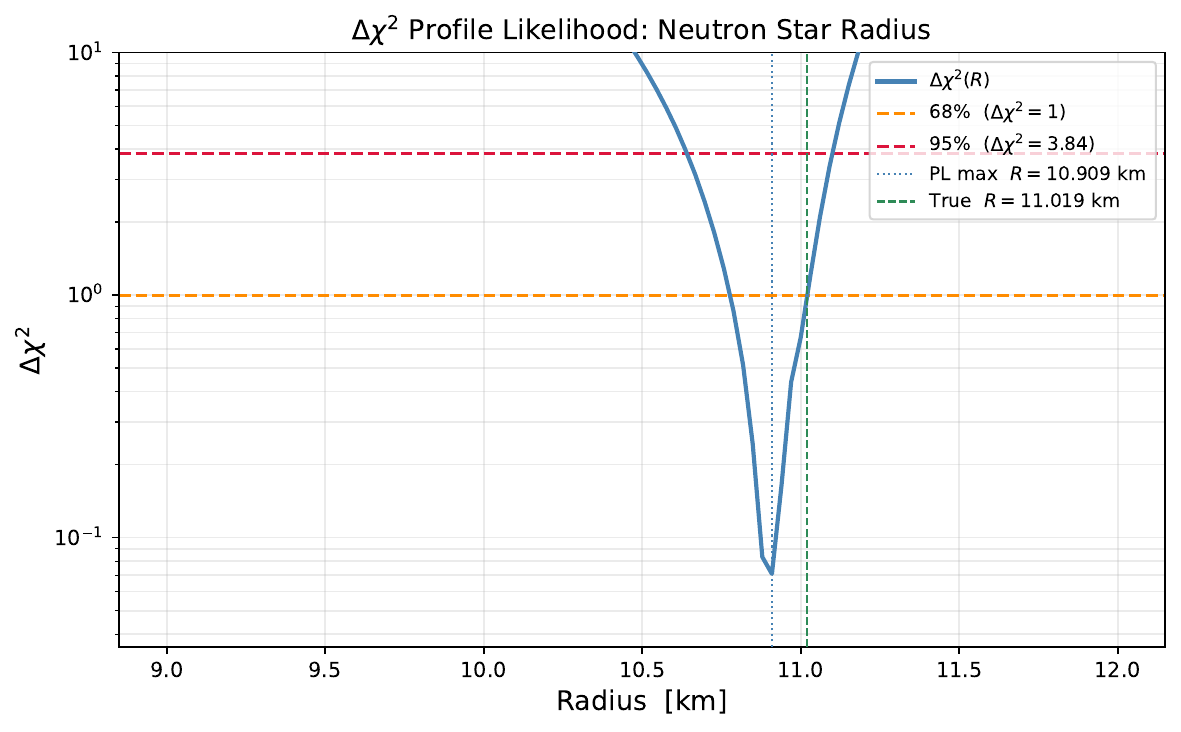}
    \caption{$\Delta\chi^2$  as a function of the neutron star radius, inferred from X-ray pulse profile modeling of synthetic data subset \textit{syntX2}, where $\Delta\chi^2 = -2(\ln\mathcal{L}_{\mathrm{profile}} - \ln\mathcal{L}_{\mathrm{max}})$ \rthis{similar to  Fig.~\ref{fig2}. The horizontal dashed lines mark the 68\% ($\Delta\chi^2 = 1$) and 95\% ($\Delta\chi^2 = 3.84$) confidence thresholds. The dotted green vertical line denotes the true injected radius, while the blue dotted line illustrates the recovered best-fit radius from the profile likelihood analysis.}}
    \label{fig:d2PL}
\end{figure}

We now perform the same analysis as in the main manuscript for an additional synthetic dataset. The second dataset  we use is described in the second column of Table 1 of H25, and is labelled \textit{syntX2}.  The injected radius for this dataset is equal to 11.019 km.
The Bayesian posterior credible intervals obtained using {\tt MultiNest} are shown in Fig.~\ref{fig:d2B}. The marginalized 68\% credible interval value for the radius is  $R=11.014^{+0.052}_{-0.055}$ km.
The corresponding profile likelihood plot is shown in Fig.~\ref{fig:d2PL}. We find that the frequentist estimate of the radius at 68\% confidence level  is    $R=10.909^{+0.110}_{-0.131}$ km. Therefore, the radius estimated using the profile likelihood agrees with the injected radius to within $1\sigma$. However, for this dataset, the Bayesian estimate is closer to the injected radius than the frequentist value.

\authorcontributions{``Conceptualization, S.D. and V.R.; methodology, V.R.; software, V.R.; validation, V.R.; formal analysis, V.R.;  writing---original draft preparation, S.D. and V.R.; writing---review and editing, S.D. and V.R.; All authors have read and agreed to the published version of the manuscript.''}


\dataavailability{No real observational data were used for this analysis. The synthetic data used for the analysis  are available at \url{https://zenodo.org/records/17776943}. The analysis codes used for this work have been made available at \url{https://github.com/vyaas3305/xpsi-frequentist-inference}. }

\acknowledgments{During the preparation of this manuscript/study, the author(s) used Anthropic Claude and OpenAI ChatGpt for help with code changes. The authors have reviewed and edited the output and take full responsibility for the content of this publication. The authors also thank Mariska Hoogkamer for her valuable assistance in resolving installation issues encountered during the setup of \texttt{X-PSI}. We are also grateful to the anonymous referees for constructive feedback on our manuscript.
Computational work was supported by the National Supercomputing Mission (NSM), Government of India, through access to the ``PARAM SEVA'' facility at IIT Hyderabad. The NSM is implemented by the Centre for Development of Advanced Computing (C-DAC) with funding from the Ministry of Electronics and Information Technology (MeitY) and the Department of Science and Technology (DST). }

\conflictsofinterest{The authors declare no conflicts of interest.}

\reftitle{References}
\bibliography{main}





\end{document}